\documentclass[pre,twocolumn,showpacs]{revtex4}
\usepackage{epsfig}

\newcommand{\g}[1]{{\bf {#1}}}

\begin{document}

\title{Toroidal bubbles with circulation in ideal hydrodynamics.
A variational approach}
\author{V.~P. Ruban$^{1,2}$}
\email{ruban@itp.ac.ru}
\author{ J.~Juul Rasmussen$^{2}$}
\email{jens.juul.rasmussen@risoe.dk}
\affiliation{$^{1}$L. D. Landau Institute for Theoretical Physics,
2 Kosygin Street, 119334 Moscow, Russia}
\affiliation{$^{2}$Optics and Fluid Dynamics Department, OFD-129,
Ris{\o} National Laboratory, DK-4000 Roskilde, Denmark}

\date{\today}

\begin{abstract}
Incompressible, inviscid, irrotational, and unsteady flows with circulation
$\Gamma$ around a distorted toroidal bubble are considered.
A general variational principle that determines the evolution
of the bubble shape is formulated. For a two-dimensional (2D) cavity 
with a constant area $A$, exact pseudo-differential
equations of motion are derived, based on variables that determine a 
conformal mapping of the unit circle exterior into the region occupied
by the fluid. A closed expression for the Hamiltonian
of the 2D system in terms of canonical variables is obtained.
Stability of a stationary drifting 2D hollow vortex is demonstrated, 
when the circulation is relatively large, $gA^{3/2}/\Gamma^2\ll 1$.
For a circulation-dominated regime of three-dimensional flows
a simplified Lagrangian is suggested, inasmuch as
the bubble shape is well described by the center-line
$\g{R}(\xi,t)$ and by an approximately circular cross-section with
relatively small area, $A(\xi,t)\ll (\oint |\g{R}'|d\xi)^2$.
In particular, a finite-dimensional dynamical system is derived and
approximately solved for a vertically moving axisymmetric vortex ring bubble
with a compressed gas inside.
\end{abstract}



\maketitle


\section{Introduction}

Vortex ring bubbles in water are like usual vortex rings with circulation,
but the core is filled with air, thus they are also termed ``air-core vortex 
rings''. The higher velocity fluid surrounding the core of the ring is at 
lower pressure than the fluid farther away due to the Bernoulli effect. 
Vortex ring bubbles can be generated in various ways naturally
or artificially, and they are interesting objects both from experimental 
and theoretical points of view. Amazing examples of the natural beauty
are vortex bubbles blown by dolphins for amusement. Also whales sometimes 
blow ring bubbles that can reach several meters in diameter. 
In laboratory conditions, toroidal bubbles can be created relatively 
easy by an air jet that is rapidly opened and closed at the bottom of a 
water tank, as in the early experiments by Walters and Davidson \cite{WD63}. 
The toroidal bubbles with circulation were formed as the result of 
gravity-induced topological transformation of an initial large spherical 
bubble, when a "tongue" of liquid penetrated the bubble from below. 
This appears to be a generic way of the creation of bubble rings 
(see, e.g.,  \cite{LM1991,SussmanSmereka}, and references therein). 
Such spherical bubbles may also be produced in nature, for instance, 
by underwater explosions. When formed the toroidal bubbles propagate 
upwards with an increasing diameter. Amusing examples of the generation 
and dynamics of vortex ring bubbles - or ``silver rings'' - may be found 
at the web site \cite{bubblerings}.

The first attempts to describe the dynamics of the vortex ring bubbles 
analytically have been made a long time ago 
(see \cite{Ponstein1959,Pedley,LM1991}, and references therein).
It is clear that the most general and realistic theoretical
consideration should be based on the Navier-Stokes equations,
and thus is a very complicated  nonlinear free-boundary problem in 
three-dimensional (3D) space. However, in many cases the inviscid
approximation, based on the Euler equations, may provide useful results.
Inviscid flows belong to the class of conservative dynamical systems 
and thus are more easily studied by Hamiltonian and Lagrangian methods
\cite{Arnold,Z68,LMMR86,DKSZ96,DLZ95,ZD96,DZK96,D2001,ZDV2002,ZK97,IL,HMR,
Salmon,Morrison98}.
With these methods, it is possible to  simplify the analysis considerably
and make it more compact, especially for irrotational flows, when the
original 3D problem  reduces  to effectively a 2D problem on the free surface
\cite{Z68,LMMR86,DKSZ96,DLZ95,ZD96,DZK96,D2001,ZDV2002,ZK97}. 

The vortex ring bubble is a special example of a general class of 
fluid dynamical problems involving the free surface separating the 
fluid and air (or generally two different fluids). The aim in this paper 
is to develop a Hamiltonian formalism for these systems, and the symmetric 
vortex ring bubble will be treated as a particular example. In other terms,
we consider the question about the
principle of least action for a general toroidal bubble.
The corresponding Lagrangian functional is shown to possess,
besides quadratic (inertial) terms on generalized velocities, 
gyroscopic terms (of the first order on generalized velocities). 
The gyroscopic terms are proportional 
to the constant circulation along linked contours.
This property makes the toroidal bubble similar to a vortex filament, 
if the circulation is large. We should emphasize that our approach, which 
is based on inviscid flows, can naturally not describe the topological 
transformation of, e.g., a raising spherical bubble into a vortex 
ring bubble as briefly discussed above.

Having obtained a general variational formulation, we derive various
approximations with reduced number of degrees of freedom. 
First of all, we consider the exact reduction corresponding to purely 2D flows
around a cavity. In this case it is possible to express the Lagrangian
in terms of the so-called conformal variables 
\cite{DKSZ96,DLZ95,ZD96,DZK96,D2001,ZDV2002}.
For 3D flows, we do not have an exact explicit expression for the Lagrangian, 
but approximations are possible. 
Such approximate dynamical systems take into account only the
most relevant degrees of freedom of the original system. 
In this way we have obtained  an approximate Lagrangian 
for a relatively long and thin toroidal bubble. 
For an axially symmetric rising and spreading vortex ring
bubble our variational approach provides a finite-dimensional approximate
system that is a generalization of the model discussed by Lundgren and 
Mansour \cite{LM1991}.

The remaining of the paper is organized as follows: 
In Sec. II we derive the general variational principle for the bubbles. 
We consider as an illustration the example of bubbles without circulation, 
before we derive the general Lagrangian for bubbles with circulation. 
The two-dimensional hollow vortex is considered in Sec. III, while the 
three-dimensional hollow vortex with a general toroidal shape is considered 
in Sec. IV. As a specific example we investigate the dynamics of the 
axisymmetric vortex ring bubble. Finally, Sec. V contains our summary.


\section{Variational principle for bubbles}
\subsection{Hamiltonian structure of equations of motion}

It is well known that a class of irrotational solutions exists in the 
framework of ideal hydrodynamics. Such solutions describe potential flows
with zero curl of the velocity field at any moment of time in the bulk
of the moving fluid. If the liquid is also incompressible 
(with the unit density, for simplicity) then the investigation
of non-stationary irrotational flows in a space region $D$ with the free
surface $\Sigma$ can be reduced to the consideration of Hamiltonian dynamics
of the surface \cite{Z68,LMMR86,DKSZ96,DLZ95,ZD96,DZK96,D2001,ZDV2002,ZK97}.
In this formulation, the shape of the surface $\Sigma$ itself
and the boundary value $\Psi$ of the velocity potential are the dynamical 
variables determining the state of the system. The velocity potential of 
incompressible fluid satisfies the Laplace equation in the bulk of the fluid
\begin{equation}\label{irrotational}
{\bf V}({\bf r},t)=\nabla \Phi,\qquad \Delta \Phi=0, 
\qquad \Phi|_\Sigma=\Psi.
\end{equation}
Besides the free surface $\Sigma$, in the general case the total boundary 
of the region $D$ has other pieces, which consist of infinitely far points 
of $D$ and/or some walls (such as surfaces of submerged bodies). 
For the remaining boundary conditions for $\Phi({\bf r})$, we will suppose
that $\Phi({\bf r})$ vanishes at infinitely far points,
and employ the no-penetration condition at a motionless wall $W$ 
(in particular, it can be the bottom with arbitrary profile), if it is present:
\begin{equation}\label{boundary_cond}
(\nabla \Phi\cdot{\bf N})|_W=0, \qquad \Phi|_\infty=0,
\end{equation}
where ${\bf N}$ is a normal vector on the wall.

The equations of motion for $\Sigma$ and $\Psi$ take the following form:
\begin{eqnarray}\label{dotSigma}
\dot\Sigma&=&V_n\equiv({\bf n}\cdot\nabla\Phi)|_\Sigma, \\
\label{dotPsi}
-\dot\Psi&=&\left(-\dot\Sigma\cdot\frac{\partial \Phi}{\partial n}
+\frac{V^2}{2}+gz+P({\cal V})\right)\Big|_\Sigma \nonumber \\
&=&\left(-V^2_n+\frac{V^2}{2}+gz+ P({\cal V})\right)\Big|_\Sigma.
\end{eqnarray}
This set of equations describe, e.g., the dynamics of a bubble or a void 
submerged into a fluid.
$\dot\Sigma$ is the speed of surface motion along the
normal unit vector ${\bf n}$ directed inside the bubble, $V_n$ is the normal 
component of the velocity field, and $\dot\Psi$ is the total time derivative 
of the boundary value of the potential $\Psi$ on the moving surface 
\cite{LMMR86,ZK97}.
The (normalized to the fluid density) pressure $P({\cal V})$ 
of the gas inside the bubble (the gas is considered as approximately massless
and adiabatic) depends on the total volume of the bubble,
\begin{equation}\label{volume}
{\cal V}=-\frac{1}{3}\int_\Sigma({\bf r}\cdot {\bf n})dS,
\end{equation}
where $dS$ is an element of the surface area.
The vertical Cartesian coordinate $z$ is measured from the horizontal plane
where the pressure is zero at the equilibrium. The gravitational acceleration 
is $-g{\bf e}_z$. Thus, at the horizontal surface of the fluid at 
the atmospheric pressure (for instance, at the see level), 
$z=z_*\approx -10$ m. Eq.(\ref{dotSigma}) is simply
the kinematic condition, and Eq.(\ref{dotPsi}) follows from the Bernoulli 
equation for non-stationary potential flows \cite{LL6}.

It is possible to verify that the right-hand sides of the equations
(\ref{dotSigma})-(\ref{dotPsi}) have the form of variational derivatives
\begin{equation}\label{Ham_form}
\dot\Sigma=\frac{\delta{\cal H}\{\Sigma,\Psi\}}{\delta \Psi},\qquad
-\dot\Psi=\frac{\delta{\cal H}\{\Sigma,\Psi\}}{\delta \Sigma},
\end{equation}
where the Hamiltonian ${\cal H}\{\Sigma,\Psi\}$ is the sum of the kinetic 
energy of the moving fluid, the internal energy of the compressed gas,
and the potential energy in the uniform gravitational field 
(all the quantities are normalized to the fluid density),
\begin{equation}\label{Ham3D}
{\cal H}=\frac{1}{2}\int_D(\nabla\Phi)^2~d{\bf r} +{\cal E}({\cal V})
+\frac{g}{2}\int_\Sigma({\bf e}_z\cdot{\bf n})z^2 ~dS.
\end{equation}
Here the adiabatic relation  between the internal energy
and the pressure is used,
\begin{equation}\label{EP}
{\cal E}'({\cal V})=- P({\cal V}).
\end{equation}
The derivation of the equality
${\delta{\cal H}}/{\delta \Psi}=V_n$ is easy. Indeed, due to
the equations (\ref{irrotational})-(\ref{boundary_cond})
one can write,
$$
\delta{\cal H}|_{\delta\Sigma=0}=
\int_D\nabla\Phi\cdot\nabla\delta\Phi~d{\bf r}=
\int_\Sigma V_n\delta\Psi~dS .
$$
The calculation of ${\delta{\cal H}}/{\delta \Sigma}$ is a bit 
more involved. It consist of two parts. First, due to the variation
$\delta\Sigma$ (in the normal direction) of the integration domain $D$ 
without changing the potential $\Phi$ inside, the following terms arise
$$
\delta{\cal H}^{(1)}|_{\delta\Psi=0}=
\int_\Sigma\left(\frac{V^2}{2}+gz+ P({\cal V})\right)\delta\Sigma~dS.
$$
The second part comes from the condition that the value $\Psi$ 
on the new $(\Sigma+\delta\Sigma)$-surface must remain the same as on the old 
$\Sigma$-surface. To satisfy this requirement, the potential $\Phi$ on the old
boundary should acquire the change $-(\partial\Phi/\partial n)\delta\Sigma$. 
Therefore the second term is
$$
\delta{\cal H}^{(2)}|_{\delta\Psi=0}=
\int_\Sigma\frac{\delta{\cal H}}{\delta \Psi}\cdot
\left(-\frac{\partial \Phi}{\partial n}\right)\delta\Sigma~dS=
-\int_\Sigma V_n^2~\delta\Sigma~dS
$$
The comparison of the sum of two these parts with the Eq.(\ref{dotPsi})
gives the second equation from Eq.(\ref{Ham_form}).

Finally, we note that surface tension  can be accounted for by simply 
adding the term $\sigma\int_\Sigma dS$, which is the surface energy, 
to the Hamiltonian (\ref{Ham3D}). Here  $\sigma$ designates the 
surface tension coefficient (divided by the fluid density).

\subsection{Variational principle}
\subsubsection{Bubbles without circulation}

It is observed that in the simplest case, when the potential $\Phi$ is a
single-valued function, the equations of motion (\ref{Ham_form}) follow from
the variational principle $\delta{\cal A}=\delta\int{\cal L}dt =0$
with the Lagrangian
\begin{equation}\label{Lag3D}
{\cal L}=\int_\Sigma\Psi\dot\Sigma~dS-{\cal H}\{\Sigma,\Psi\}.
\end{equation}
This expression is written in the invariant form that does not depend on the
choice of the parameterization for the surface shape. Practically,
this choice is dictated by geometry of a given problem.
For instance, the parameterization $z=\eta(x,y,t)$
is commonly used to study  waves on the sea surface. It is clear
that due to the equality $\Psi\dot\Sigma ~dS=\Psi\eta_t ~dxdy$ the functions
$\eta(x,y)$ and $\Psi(x,y)$ form a pair of canonically conjugated 
variables \cite{Z68,ZK97}.  But if we want to study oscillations 
of a spherical bubble, the spherical coordinates $r,\theta,\varphi$ 
are more convenient. In this case the functions $-\Psi(\theta,\varphi)$ and
$Q(\theta,\varphi)=r^3(\theta,\varphi)/3$ can be taken as canonical variables.

As an illustrative example, we consider the case corresponding to
spherically symmetric flows with $g=0$ and with a constant external 
pressure $P_{ext}$. In this case the dynamical variables depend only on $t$,
and we have the completely solvable conservative system for
${\Psi}(t)$ and ${\cal V}(t)$, represented by the Lagrangian:
\begin{equation}\label{sonolum}
{\cal L}_{\mbox{\scriptsize sph.}}=
-{\Psi}\dot{\cal V}-a{\cal V}^{1/3}\frac{{\Psi}^2}{2}
-{\cal E}({\cal V})-P_{ext}{\cal V} -b{\cal V}^{2/3},
\end{equation}
where $a=3^{1/3}(4\pi)^{2/3}$, and $b=3^{2/3}(4\pi)^{1/3}\sigma$ 
accounts for the surface tension. The equations of motion 
-- the Euler-Lagrange equations -- corresponding to this Lagrangian are: 
$$
-\dot{\cal V}-a{\cal V}^{1/3}\Psi=0,
$$
$$
\dot\Psi-\frac{a}{6{\cal V}^{2/3}}{\Psi^2}-{\cal E}'({\cal V})-P_{ext}
-\frac{2}{3} \frac{b}{{\cal V}^{1/3}} = 0.
$$
From here one can eliminate $\Psi$ and obtain the equation of the
second order for ${\cal V}$:
\begin{equation}
\frac{\ddot{\cal V}}{a {\cal V}^{1/3}} - 
\frac{1}{6} \frac{\dot{\cal V}^{2}}{a {\cal V}^{4/3}} 
+ P_{ext} + {\cal E}' ({\cal V}) + 
\frac{2}{3} \frac{b}{{\cal V}^{1/3}} = 0.
\label{sonovol}
\end{equation}
It is easy to show that Eq.(\ref{sonovol}) is equivalent 
to the simplest variant of the Rayleigh-Plesset equation for the radius of 
a spherical bubble (see \cite{HBGL1998} and references therein), 
using $ {\cal V} = ({4 \pi}/{3}) R^3 $,
$$
R \ddot{R} + \frac{3}{2} \dot{R}^2 = - P_{ext} - {\cal E}'({\cal V}) 
 - \frac{2 \sigma }{R}.
$$
Since the Lagrangian (\ref{sonolum}) does not depend explicitly on time, 
the system possesses the energy integral
$$
a{\cal V}^{1/3}\frac{{\Psi}^2}{2}
+{\cal E}({\cal V})+P_{ext}{\cal V} +b{\cal V}^{2/3}=E_0.
$$ 
Therefore the solution of the Eq.(\ref{sonovol}) is determined by 
$$
t=\int_{{\cal V}_0}^{\cal V}\! \!\frac{d\tilde{\cal V}}
{\sqrt{2a\tilde{\cal V}^{1/3}[E_0-{\cal E}(\tilde{\cal V})-
P_{ext}\tilde{\cal V}-b\tilde{\cal V}^{2/3}]}},
$$
where $E_0$ and ${\cal V}_0$ are arbitrary constants.
If ${\cal E}({\cal V})=0$ (the bubble may contain no gas),
then the above expression describes a spherical cavity collapse for 
$P_{ext}>0$, as well as possible cavity formation for negative $P_{ext}$.

More complex spherical bubble dynamics with a time-dependent $P_{ext}(t)$
is governed by the Lagrangian (\ref{sonolum}) as well, however, we do not 
have analytical solutions for that case. For instance, the dependence 
$P_{ext}(t)=P_0+P_{s}\cos(\omega t)$ is related to the problem of single 
bubble sonoluminescence \cite{HBGL1998}, where $\omega$ is the frequency
of a (relatively long) standing ultrasound wave.

\subsubsection{Toroidal bubbles with circulation}

The variational formulation becomes more complicated in the case 
when the free surface $\Sigma:(\vartheta,\xi)\mapsto{\bf R}=(X,Y,Z)$, with
$0\le\vartheta<2\pi$ and $0\le\xi<2\pi$,
is topologically equivalent to a torus, and the circulation of the velocity
along linked contours takes a nontrivial value $\Gamma$.
Now the potential $\Phi$ is a multi-valued function,
\begin{equation}
\Phi=\phi+(\Gamma/2\pi)\theta,
\end{equation}
where $\phi$ is the harmonic potential determined by a single-valued
boundary  function $\psi(\vartheta,\xi)$,
and the velocity field created by the multi-valued
harmonic function $\theta$ has zero normal component on the free surface.
The  important point is that the potential $\theta$ is completely determined
by the shape of the toroidal bubble. The multi-valued  boundary function
$\Theta(\vartheta,\xi)$ associated with
the potential $\theta$ increases by the value $2\pi$ as the coordinate
$\vartheta$ acquires the increase $2\pi$.
The kinetic energy of the flow is represented as the sum
of circulation-induced energy and the energy associated with
the motion of the bubble. In the general form, we have the following expression:
\begin{equation}\label{kinetic_energy}
{\cal K}=\Gamma^2{\cal K}_{\mbox{\scriptsize c}}\{\Sigma\}+
\frac{1}{2}\int\!\!\int
 {\cal G}_{\Sigma}(s_1,s_2)\psi(s_1)\psi(s_2) dS_1dS_2,
\end{equation}
where $s_1\in\Sigma$, $s_2\in\Sigma$, and ${\cal G}_{\Sigma}(s_1,s_2)$
is a symmetric function completely determined by a given shape of the bubble.

In order to have correct equations of motion for ${\bf R}(\vartheta,\xi,t)$ 
and $\psi(\vartheta,\xi,t)$ (the equations must be equivalent to 
Eq.(\ref{Ham_form})), 
it is necessary to include into the action ${\cal A}=\int{\cal L}dt$ a term 
that will give the same contribution as the following term,
$$
\frac{\Gamma}{2\pi}\int\! dt\!\int_\Sigma\Theta\dot\Sigma dS
= \frac{\Gamma}{2\pi}\int\!\!\int\!\!\int
([{\bf R}_{\xi}\times{\bf R}_{\vartheta}]\cdot{\bf R}_t)\Theta
~dt~d\vartheta~d\xi.
$$
It is clear that this expression should be transformed by some integration 
in parts to a form where $\Theta$ is not employed, but only the derivatives
$\Theta_t$, $\Theta_\xi$, and $\Theta_\vartheta$
that are single-valued functions. As a result, we obtain
that the Lagrangian for a hollow vortex tube can be written as follows
\begin{eqnarray}
&&{\cal L}=\int_\Sigma\psi\dot\Sigma~dS -{\cal H}\{\Sigma,\psi\}\nonumber\\
&&\quad+\frac{\Gamma}{3\cdot 2\pi}
\int {\bf R}\cdot\Big\{
[{\bf R}_{\xi}\times{\bf R}_t]\Theta_{\vartheta}\nonumber\\
\label{hollow3D}
&&\qquad+[{\bf R}_t\times{\bf R}_{\vartheta}]\Theta_{\xi}
-[{\bf R}_{\xi}\times{\bf R}_{\vartheta}]\Theta_t\Big\}
~d\vartheta ~ d\xi.
\end{eqnarray}
Now we may identify the function $\Theta$ with the coordinate $\vartheta$
and thus the two last terms are equal to zero.
Also it is possible in general to express the potential $\psi$ as
\begin{equation}\label{psi}
\psi(s)=\int {\cal M}_{\Sigma}(s,\tilde s)\dot\Sigma(\tilde s) d\tilde S,
\end{equation}
where the ``matrix'' ${\cal M}_{\Sigma}$ is the inverse of the ``matrix''
${\cal G}_{\Sigma}$, and thus exclude $\psi$ from the Lagrangian.
Then we will obtain the Lagrangian of the form
\begin{eqnarray}
&&{\cal L}=\frac{\Gamma}{3\cdot 2\pi}\int
({\bf R}\cdot[{\bf R}_{\xi}\times{\bf R}_t])~d\vartheta ~ d\xi
-\Pi\{\Sigma\}\nonumber\\
&&\qquad+\frac{1}{2}\int\!\! \int {\cal M}_{\Sigma}(s_1,s_2)
\dot\Sigma(s_1)\dot\Sigma(s_2) dS_1dS_2,
\label{LagrangianSigma}
\end{eqnarray}
where the effective potential energy $\Pi\{\Sigma\}$ is the sum of
the circulation-induced energy, the internal energy
of the compressed gas inside the bubble, the gravitational energy of the bubble,
and the surface energy,
\begin{equation}\label{Pi}
\Pi\{\Sigma\}=\Gamma^2{\cal K}_{\mbox{\scriptsize c}}\{\Sigma\}
+{\cal E}({\cal V})+\frac{g}{2}\int_\Sigma({\bf e}_z\cdot{\bf n})z^2 ~dS 
+ \sigma\int_\Sigma dS.
\end{equation}

It is interesting to note that for circulation-dominated configurations
(it is important that the gradient of $\Psi$ along the surface
should not be equal to zero at any point of $\Sigma$),
a similarity exists between a hollow vortex tube and 
an ordinary toroidal vortex sheet. Indeed, the dynamics 
of a toroidal vortex sheet in a fluid without free boundary
is governed by the Lagrangian (see, for instance, \cite{R2001PRE}
and references therein)
\begin{equation}\label{sheet3D}
{\cal L}_{\mbox{\scriptsize v.sh.}}=\frac{1}{3}
\int([{\bf R}_{\xi}\times{\bf R}_t]\cdot{\bf R})~d\nu ~ d\xi 
-{\cal H}_{\mbox{\scriptsize v.sh.}}\{{\bf R}(\nu,\xi)\},
\end{equation}
where the vector function ${\bf R}(\nu,\xi)$ describes the shapes of
individual vortex lines enumerated by the label $\nu\in[0,\Gamma]$, 
and ${\bf R}_{\xi}$ is directed along the vortex lines on the sheet.
On the other hand, when considering the hollow vortex tube
with a strong circulation, we could use the function $\Psi$ 
as a coordinate on the bubble surface, instead of the coordinate 
$\vartheta$, and in that case the Lagrangian of the hollow vortex 
tube would take the alternative form:
\begin{equation}\label{L_hollow_alternative}
{\cal L}=\frac{1}{3}
\int([{\bf R}_{\xi}\times{\bf R}_t]\cdot{\bf R})~d\Psi ~ d\xi 
-{\cal H}\{{\bf R}(\Psi,\xi)\},
\end{equation}
where ${\cal H}\{{\bf R}(\Psi,\xi)\}$ is the total energy of the toroidal 
bubble. Thus, the only difference between Eq.(\ref{sheet3D}) and
Eq.(\ref{L_hollow_alternative}) is in the Hamiltonians 
${\cal H}_{\mbox{\scriptsize v.sh.}}$ and ${\cal H}$. 
In the limit of a ``thin vortex tube'' the Hamiltonians are almost identical,
inasmuch as the main contribution is due to the logarithmically
large circulation-induced kinetic energy.

In the general case a free surface may consist of several separated manifolds
with nontrivial topology. All of these must be included into the Lagrangian
in a similar manner.


\section{2D hollow vortex}

As application of the theory described in the previous section,
let us first consider a 2D irrotational flow in the $yz$-plane, with the 
circulation $\Gamma=2\pi\gamma$ around a cavity having a finite area 
$A=\pi r^2$. The 2D geometry allows us to employ the theory of conformal 
mappings to derive exact equations of motion for such a system. 
Conformal variables have been extensively used during recent years 
for analytical studies of waves on water surface, 
and for numerical simulations 
(see, for instance \cite{DKSZ96,DLZ95,ZD96,DZK96,D2001,ZDV2002}).
The system considered in this section has a set of additional 
properties in comparison with the usual surface waves. The
presence of the circulation makes it similar to a vortex. At the same time,
the hollow vortex possesses inertial properties and a potential energy
in the gravitational field.  For small values of the parameter
$\mu=gr^3/\gamma^2$ a stationary horizontal drift of the hollow vortex 
is possible with the velocity $V_d\approx-gr^2/(2\gamma)$ and with the shape 
close to circular. This motion is stable, as will be discussed below.
Therefore the content of this section will serve as a basis for further
simplified descriptions of 3D circulation-dominated flows.

\subsection{Conformal mapping}

We consider an infinite two-dimensional region $D$, which is topologically
equivalent to the exterior of the unit circle. Our purpose is to obtain
an expression for the kinetic energy of the irrotational flow with the
circulation $2\pi\gamma$ around the cavity in the case of an
arbitrary given shape of the surface and arbitrary given boundary potential
$\Psi$ (with the only condition $\Psi\mapsto\Psi+2\pi\gamma$ after 
one turn along the boundary.
Strictly speaking, this energy is infinite because of the divergency of the 
corresponding integral at the infinity. But this is not important for the
equations of motion, inasmuch as the presence of an infinite constant term 
in the Hamiltonian in no way influences the dynamics. Therefore only the excess
of the energy in comparison with some basic state is needed. As the basic state,
we shall take the perfect circular shape of the boundary, with the radius $r$
and purely azimuthal velocity field, inversely proportional to the distance
from the central point.

Since the velocity potential, $\Phi(y,z,t)$, satisfies the Laplace equation
$\Phi_{yy}+\Phi_{zz}=0$,
which is conformally invariant, it is natural to re-formulate the problem
in terms of the conformal mapping of the unit circle exterior into the region 
$D$. This mapping is determined by an analytical function 
$\zeta_*(w,t)=y(w,t)+iz(w,t)$ of a complex variable $w$. 
The function $\zeta_*(w,t)$ has no
singularities at $|w|>1$ and behaves proportionally to $w$ when $w\to\infty$.
Therefore the expansion of this function in powers of $w$ contains no positive 
powers higher than $1$. The shape of the free surface is given parametrically
by the expression
\begin{eqnarray}
&&Y(\vartheta,t)+iZ(\vartheta,t)=
\zeta_*(w,t)\big|_{w=e^{i\vartheta}}\nonumber\\
&&\equiv\zeta(\vartheta,t)=\zeta_1(t)e^{i\vartheta}
+\sum_{m=-\infty}^{0}\zeta_{m}(t) e^{im\vartheta}.
\end{eqnarray}
In the potential $\Psi$ we now explicitly separate  the term
$\gamma\vartheta$, which is responsible for the circulation,
\begin{equation}
\Psi(\vartheta,t)=\gamma\vartheta+\psi(\vartheta,t),
\end{equation}
\begin{equation}
\psi(\vartheta,t)=\sum_{m=-\infty}^{+\infty}\psi_{m}(t) e^{im\vartheta},
\qquad \psi_{-m}=\bar \psi_m.
\end{equation}
The term $\gamma\vartheta$ corresponds to the multi-valued harmonic function
$\Phi_0(w)=\mbox{Re}(-i\gamma\,\mbox{Ln\,} w)$ with zero normal component 
of the velocity at the free surface. The single-valued function $\psi$ 
is related to evolution of the boundary shape. It can be understood as the
potential of surface waves. The excess of energy is the sum of two parts. 
The first part is due to the kinetic energy of the surface waves,
$$
E_{\mbox{\scriptsize s.w.}}=
2\pi\sum _{m=-\infty}^{+\infty}\frac{|m||\psi_m|^2}{2}.
$$
The other part arises in the circulational energy as the result of changing of
the effective cavity size, and it is completely determined by the coefficient
$\zeta_1$:
$$
E_\gamma=-\frac{2\pi\gamma^2}{4}\ln\Big|\frac{1}{r}
\int \zeta(\vartheta)e^{-i\vartheta}\frac{d\vartheta}{2\pi}\Big|^2.
$$

Now we have to introduce some necessary linear operators
\cite{DKSZ96,DLZ95,ZD96,DZK96,D2001,ZDV2002}
to deal with boundary values of analytical functions.
In Fourier-representation these operators are diagonal,
\begin{equation}
\hat H_m=i~\mbox{sign}(m),\quad
\hat M_m=|m|,\quad
\hat P^{(\mp)}_m=\frac{1}{2}(1\mp \mbox{sign}(m)).
\end{equation}
Here the operator $\hat H$ is the Hilbert transformation.
The operator $\hat P^{(-)}$ excludes the Fourier-harmonics with positive $m$, 
while $\hat P^{(+)}$ excludes the harmonics with negative $m$.
The following equalities will be used in the further exposition,
\begin{equation}
\hat M=-\hat H \partial_\vartheta,\quad
\hat P^{(\mp)}=\frac{1}{2}(1\pm i\hat H),\quad
\hat P^{(+)}+\hat P^{(-)}=1.
\end{equation}

We have now prepared all the necessary tools, and we are able to write down 
the Lagrangian for a 2D hollow vortex in the conformal variables:
\begin{eqnarray}
&&{\cal L}_{\mbox {\scriptsize conf.}}=
-\gamma\int\frac{(\dot \zeta\bar \zeta-\dot{\bar \zeta}\zeta)}{4i}
d\vartheta
+\int\psi\frac{(\dot \zeta\bar \zeta'-\dot{\bar \zeta}\zeta')}{2i}
d\vartheta\nonumber\\
&&+\frac{2\pi\gamma^2}{4}
\ln\Big|\frac{1}{r}\int \zeta(\vartheta)e^{-i\vartheta}
\frac{d\vartheta}{2\pi}\Big|^2
-\frac{\gamma^2}{2r^2}\int\frac{( \zeta'\bar \zeta-\zeta\bar \zeta')}
{4i}d\vartheta\nonumber\\
&&-\frac{1}{2}\int\psi\hat M \psi ~d\vartheta
-\frac{g}{2}\int\left(\frac{\zeta-\bar \zeta}{2i}\right)^2
\frac{(\zeta'+\bar \zeta')}{2}d\vartheta \nonumber\\
\label{L_z_psi}
&&+\int\left(\bar\lambda\hat P^{(+)}(\zeta e^{-2i\vartheta})
       +\lambda\hat P^{(-)}(\bar \zeta e^{2i\vartheta})\right)d\vartheta.
\end{eqnarray}
Here $\dot \zeta=\partial_t \zeta$, $\zeta'=\partial_\vartheta \zeta$.
Besides the obvious terms that were already explained in the previous 
discussion, in the Lagrangian ${\cal L}_{\mbox {\scriptsize conf.}}$
there is the term proportional to the 
constant area of the cavity. Its presence provides 
minimum of the circulational part of the Hamiltonian on the perfect shape 
$\zeta=re^{i\vartheta}+\zeta_0$. To be punctual, we have also included the 
terms with the Lagrangian multipliers $\lambda$ and $\bar\lambda$ in order 
to specify explicitly the analytical properties of the function 
$\zeta(\vartheta)$.

The variation of the action with the Lagrangian (\ref{L_z_psi}) gives 
(after some additional transformations, see Appendix A) the equations of
motion for $\zeta(\vartheta,t)$ and $\psi(\vartheta,t)$,
\begin{eqnarray}\label{dot_z}
\dot \zeta&=&\zeta'\hat P^{(-)}\left(\frac{2i\hat M \psi}
{|\zeta'|^2}\right),\\
\label{dot_psi}
\dot\psi&=&\frac{(\hat H\psi')^2-(\gamma+\psi')^2}{2|\zeta'|^2}+
(\gamma+\psi')\hat H\left(\frac{\hat H\psi'}{|\zeta'|^2}\right)
\nonumber \\
&&
-g\frac{(\zeta-\bar \zeta)}{2i}+\frac{\gamma^2}{2r^2}.
\end{eqnarray}
Of course, these equations can also be obtained directly by simply presenting 
the kinematic condition and the Bernoulli equation in conformal variables.

\subsection{Canonical variables}

The Lagrangian (\ref{L_z_psi}) is written in terms of variables that are not
canonically conjugated. For general purposes, e.g., of 
convenience for a nonlinear analysis, a pair
of canonical variables can be found. As the canonical coordinate,
we take the real function $q(\vartheta,t)$ such that
\begin{equation}
\zeta(\vartheta,t)=\beta(\vartheta,t)e^{i\vartheta},
\qquad \beta(\vartheta,t)=(1+i\hat H)q(\vartheta,t)
\end{equation}
After substitution into the Lagrangian (\ref{L_z_psi}), one can immediately
obtain the expression for the corresponding canonical momentum $p(\vartheta,t)$,
\begin{equation}
p=\gamma\hat H q-\hat P^{(-)}[\psi(\beta-i\beta')]
-\hat P^{(+)}[\psi(\bar \beta+i\bar \beta')]
\end{equation}
Now it is necessary to solve this equation for the potential $\psi$
in order to express the Hamiltonian in terms of $q$ and $p$.
The result of the calculations is (see Appendix B):
\begin{equation}\label{psi_pq}
-\psi\{q,p\}=\frac{(p\!-\!\gamma\hat H q)(q\!-\!\hat M q)+
\hat H[(p\!-\!\gamma\hat H q)\hat H(q\!-\!\hat M q)]}
{(q-\hat M q)^2+(\hat H(q-\hat M q))^2}.
\end{equation}
Thus, the Hamiltonian for a 2D hollow vortex is:
\begin{eqnarray}
&&{\cal H}\{q,p\}=\frac{1}{2}\int\psi\{q,p\}\hat M\psi\{q,p\}d\vartheta
\nonumber\\
&&+\frac{\gamma^2}{2r^2}\int q(1-\hat M)q ~ d\vartheta
-\frac{2\pi\gamma^2}{2}\left(\ln\left(\frac{q_0}{r}\right)
+\frac{q_0^2}{2r^2}\right)\nonumber\\
&&+\frac{g}{2}\int(q\,\sin\vartheta +\hat H q\,\cos\vartheta )^2\nonumber\\
&&\qquad\times
[(q'-\hat H q)\cos\vartheta-(q+\hat H q')\sin\vartheta]~ d\vartheta,
\end{eqnarray}
where $\psi\{q,p\}$ should be taken from Eq.(\ref{psi_pq}),
and $q_0$ is the 0-th Fourier-harmonic of the function $q(\vartheta)$,
$$
q_0=\int q(\vartheta)\frac{d\vartheta}{2\pi}.
$$

\subsection{Linearized equations for small deviations from a circular shape}

Let us consider an initial stage of evolution of a hollow vortex, 
starting from a nearly perfect -- circular -- shape (with the mean radius $r$), 
in the case of a small initial potential $\psi$. It should be emphasized that
the circular shape is not a stationary solution in the presence of gravity.
Therefore we are studying dynamics that can be, generally speaking, far 
from equilibrium. However, at least at sufficiently small times we will have
\begin{equation}
\zeta=re^{i\vartheta}b(\vartheta,t),\quad b(\vartheta,t)
=b_0+\sum_{m=1}^{+\infty}b_{-m}e^{-i\vartheta},\quad b_0\approx 1,
\end{equation}
and $|b_{-m}|\ll 1 $ if $ m>1$.
From Eq.(\ref{dot_z}) and Eq.(\ref{dot_psi}) we obtain the linearized 
(nonhomogeneous!) system
\begin{equation}
r^2\dot b_{-m}=-2m\psi_{-m}
\end{equation}
\begin{equation}
\dot \psi_{-m}=2i\frac{\gamma}{r^2} ~m\psi_{-m}+
\frac{\gamma^2}{2r^2}(1-m)b_{-m}
-\frac{gr}{2i}(b_{-(m+1)}-\delta_{m1})
\end{equation}
where $\delta_{m1} $ is the Kronecker $ \delta $.
Eliminating $\psi$, we have the set of equations
\begin{equation}
\ddot b_{-m}=2i\frac{\gamma}{r^2}m \dot b_{-m}+
\frac{\gamma^2}{r^4}(m^2-m)b_{-m}
+\frac{g}{ir}~m(b_{-(m+1)}-\delta_{m1})
\end{equation}
A particular solution of this nonhomogeneous system is
\begin{eqnarray}
\label{drift}
&&b_{-1}=\frac{gr^3}{4i\gamma^2}
\left(\exp(2i\gamma t/r^2)-1-2i\gamma t/r^2\right),\\
\label{No_distortions}
&&b_{-m}=0, \qquad m>1
\end{eqnarray}
Applicability of the linearized equations implies that the velocity 
of the vortex motion is small in comparison
with the velocity of rotation $\gamma/r$.
Thus, the parameter $\mu=gr^3/\gamma^2$ should be small, or at least
the time $t$ should be small,
$$
\mu|\sin(\gamma t/r^2)|\ll 1.
$$
If $\mu\ll 1$, then Eqs.(\ref{drift})-(\ref{No_distortions}) are 
approximately valid for arbitrary $t$ and describe the horizontal drift 
of the vortex with the mean velocity $V_d=-{gr^2}/({2\gamma})$.
The shape of the vortex in this limit remains almost circular.

The general solution of the corresponding homogeneous system is the linear
combination
\begin{equation}
b_{-m}(t)=\sum_{\nu}C_{\nu} b^{(\nu)}_{-m}
\exp\left({-it\frac{\gamma}{r^2}\Omega^{(\nu)}}\right),
\end{equation}
where $\nu$ is a discrete parameter. The dimensionless frequencies 
$\Omega^{(\nu)}$ and the corresponding modes $b^{(\nu)}_{-m}$ 
should be determined from the algebraic system
\begin{equation}\label{recurrence}
\left(m-(\Omega^{(\nu)}+m)^2\right) b^{(\nu)}_{-m}=
\frac{gr^3}{i\gamma^2}~mb^{(\nu)}_{-(m+1)}.
\end{equation}
It is clear that in the case $g\not=0$ the modes $b^{(\nu)}_{-m}$
are delocalized both in $\vartheta$-space and in $m$-representation.
They can be partially classified by the number
$n$ of the last non-zero Fourier-harmonics. Therefore $\nu=(n,...)$, and
\begin{equation}\label{Omega_n}
\left(n-(\Omega^{(n,...)}+n)^2\right) b^{(n,...)}_{-n}=0. 
\end{equation}
From here we have $\nu=(n,\pm)$, and 
\begin{equation}\label{Omega_pm}
\Omega^{(n,\pm)}=-n\pm\sqrt{n}.
\end{equation}

We see that regardless of the vortex size $r$ and the value of $\gamma$,
all the frequencies are real. From the other hand, we naturally expect
an instability for sufficiently large $r$ and/or small $\gamma$.
But there is no contradiction at this point because for a large size
and/or a small circulation the behavior of a coherent superposition
of $m$-delocalized modes with real frequencies is effectively exponential 
at small $t$. Therefore the linearized equations for small deviations 
(from the circular, not from an unknown stationary shape!) become un-valid 
very soon and the nonlinearity begins to play an essential role. 
It should be emphasized that in this subsection we have not considered 
an exact stationary configuration of the hollow vortex, since we do not 
have an analytical expression for such solution. One can expect that
a stationary shape strongly deviates from circular when the parameter
$\mu$ increases considerably, and finally, above some critical
value $\mu_*\sim 1$, a stable stationary solution does not exist. However,
at small $\mu$ stable stationary solutions do exist, 
and the stationary shape is almost circular, inasmuch as 
the linearized equations for small deviations remains approximately
valid for arbitrary time and their solutions are stable. 
The questions of the determination of the critical value $\mu_*$
and how instability develops are left for future investigations.


\section{3D hollow vortex tube}
\subsection{Simplified Lagrangian}

We proceed to a simplified consideration of a 3D thin and long
closed hollow vortex tube with a smooth center-line 
$$\g{R}(\xi,t)=(X(\xi,t),Y(\xi,t),Z(\xi,t))$$ 
and with approximately circular cross-section having a relatively small area
$A(\xi,t)=\pi a^2(\xi,t)\ll\Lambda^2$, 
where $a$ is the radius of the cross-section, 
and $\Lambda$ is the total length of the center-line,
\begin{equation}
\Lambda=\oint |\g{R}'|d\xi.
\end{equation}
This description should be good in most cases, since for large enough 
circulation ($\Gamma^2\gg g A^{3/2}$) the local
quasi-2D dynamics is stable with approximately circular cross-section,
and also for a straight thin 3D tube ($\Gamma^2>4\pi^2 \sigma (A/\pi)^{1/2}$,
see \cite{Ponstein1959}) it is easy to demonstrate stability of longitudinal 
sausage-like perturbations.

Assuming slow variation of $A(\xi,t)$ along the curve $\g{R}(\xi,t)$ and 
neglecting small distortions of the shape of the circular cross-section, we can 
give an explicit form of all terms in Eqs.(\ref{LagrangianSigma})-(\ref{Pi}).
As the result, a simplified Lagrangian  with   logarithmic 
accuracy can be written as follows,
\begin{eqnarray}
&&\tilde{\cal L}=\frac{1}{8\pi}\oint\ln\Big(\frac{\Lambda^2}{A}\Big)
(\dot A_\perp)^2|{\bf R}_{\xi}|d\xi
+\oint \frac{|\dot{\bf R}_\perp|^2}{2}A |{\bf R}_{\xi}|d\xi\nonumber\\
&&\quad+\frac{\Gamma}{3}\oint
({\bf R}\cdot[{\bf R}_{\xi}\times{\bf R}_t]) ~ d\xi
-\frac{\Gamma^2}{8\pi}\oint\ln\Big(\frac{\Lambda^2}{A}\Big)
|{\bf R}_{\xi}|d\xi\nonumber\\
&&\quad-{\cal E}\Big(\oint A |{\bf R}_{\xi}|d\xi\Big)
+g\oint Z A |{\bf R}_{\xi}|d\xi\nonumber\\
&&\quad-2\pi^{1/2}\sigma\oint A^{1/2} |{\bf R}_{\xi}|d\xi,
\label{LagrangianRA}
\end{eqnarray}
where
\begin{eqnarray}
&&\dot A_\perp=A_t-A_\xi({\bf R}_t \cdot {\bf R}_{\xi})/|{\bf R}_{\xi}|^2,\\
&&\dot {\bf R}_\perp={\bf R}_t-
{\bf R} _\xi({\bf R}_t \cdot {\bf R}_{\xi})/|{\bf R}_{\xi}|^2.
\end{eqnarray}

This Lagrangian includes the most principal inertial effects that
correspond to the dynamics of $\g{R}(\xi,t)$ and $A(\xi,t)$
on scales of the order of  $\Lambda$.
The interplay between the second order time-derivative (inertial) terms
and the circulation-originated first order terms  will result in oscillations
that are relatively fast if $\Gamma$ is large. 
However, in the circulation-dominated regime this system has interesting solutions 
with these oscillations almost not excited. Approximately such non-oscillating
solutions are determined by the Lagrangian without the inertial terms.
That means we have to find a minimum of the effective potential energy
$\Pi\{\g{R}(\xi),A(\xi)\}$ over $A(\xi)$ with fixed $\g{R}(\xi)$ and then
substitute the minimum-providing configuration $A_*(\xi)$ into  $\Pi$.
The extremal configuration $A_*(\xi)$ is determined by the following coupled
equations,
\begin{equation}\label{eq_A_equilibrium}
\frac{\Gamma^2}{8\pi A_*(\xi)}+P({\cal V})+gZ(\xi)-
\pi^{1/2}\sigma A_*^{-1/2}(\xi)=0,
\end{equation}
\begin{equation}\label{Volume_condition}
\oint  A_*(\xi) |{\bf R}'(\xi)|d\xi={\cal V},
\end{equation}
as observed from the Lagrangian (\ref{LagrangianRA}). At this point 
we meet a technical difficulty, since although Eq.(\ref{eq_A_equilibrium})
has the explicit solution
\begin{equation}\label{A_equilibrium}
A_*^{-1/2}=\frac{4\pi}{\Gamma^2}
\left(\pi^{1/2}\sigma+\sqrt{\pi\sigma^2-
({\Gamma^2}/{2\pi})[P({\cal V})+g Z(\xi)]}\right),
\end{equation}
unfortunately Eq.(\ref{Volume_condition}) for ${\cal V}$, with this expression 
for $A_*(\xi)$, is hard to solve exactly, except for the simplest case,
$P({\cal V})=0$, when $A_*(\xi)$ is not depending on the volume.
Nevertheless, approximate methods may be used in many cases and
corresponding approximate expressions for the effective Hamiltonian
${\cal H}_*\{\g{R}(\xi)\}=\Pi\{\g{R}(\xi),A_*(\xi)\}$ of the bubble center-line
can be obtained. The equation of motion will then have the general structure 
as follows,
\begin{equation}\label{dot_R}
[\g{R}_\xi\times\g{R}_t]=
\frac{1}{\Gamma}\,\frac{\delta{\cal H}_*}{\delta\g{R}},
\end{equation}
which are similar to the equations of motion for a slender vortex filament 
in a fluid without bubbles \cite{R2001PRE,R2000PRD},
however, with another Hamiltonian.
 
For a hollow vortex tube without gas inside, when $P({\cal V})=0$,
the dynamics of the center-line is described by the effective Hamiltonian
$$
{\cal H}^{P=0}_*\{\g{R}(\xi)\}=\Gamma\oint F(Z(\xi))|{\bf R}'(\xi)|d\xi,
$$
where the function $F(Z)$ is defined as follows,
\begin{equation}
F(Z)=\frac{\Gamma}{4\pi}\Big\{
\ln\Big[\tilde\Lambda\Big(C+\sqrt{C^2-Z}\Big)\Big]
+\frac{C}{C+\sqrt{C^2-Z}}\Big\}.
\end{equation}
Here $\tilde\Lambda\sim\Lambda g^{1/2}\Gamma^{-1}$ may be considered as 
approximately constant, and $C^2=2\pi^2\sigma^2g^{-1}\Gamma^{-2}$. 
The equation of motion (\ref{dot_R}) for this case can be rewritten as
\begin{equation}
\g{R}_t= F'(Z)[\g{e}_z\times \g{t}]+F(Z)\kappa\g{b},
\end{equation}
where $\g{t}$, $\g{b}$, and $\kappa$ are the unit tangent vector on the center-line,
the binormal vector, and the curvature of the line, respectively.
This equation is a generalization of the well known LIA (Localized Induction Approximation) equation
\cite{Hasimoto,R2001PRE,R2000PRD}.

\subsection{Axisymmetric motion}

An obvious application of the Lagrangian (\ref{LagrangianRA}) is for 
vertically rising and spreading axisymmetric vortex ring bubbles \cite{LM1991}.
We need only three degrees of freedom to describe this dynamics, namely the
vertical coordinate $Z(t)$, the radius of the ring $R(t)=(X^2+Y^2)^{1/2}$, 
and the total volume of the bubble ${\cal V}(t)=2\pi^2 a^2(t)R(t)$.
The corresponding finite-dimensional dynamical system is determined by the
following Lagrangian,
\begin{eqnarray}
&&L_{ZR{\cal V}}
=\frac{(\dot R {\cal V} - R \dot{\cal V})^2 }{16\pi^2R^3}
\ln\Big(C^{(A)}_{\mbox{\scriptsize log.}}\frac{R^3}{\cal V}\Big) +
\frac{\cal V}{2}(\dot R^2 + \dot Z^2)\nonumber\\
&&\qquad\qquad -2\pi\Gamma Z R\dot R -
\frac{\Gamma^2 R}{4}
\ln\Big(C^{(\Gamma)}_{\mbox{\scriptsize log.}}\frac{R^3}{\cal V}\Big)
\nonumber\\
&&\qquad\qquad-{\cal E}({\cal V})+ g Z {\cal V}-
2^{3/2}\pi\sigma R^{1/2}{\cal V}^{1/2},
\label{LagrangianZRV}
\end{eqnarray}
where the constant coefficients 
$$C^{(A)}_{\mbox{\scriptsize log.}}=128\pi^2, 
\qquad 
C^{(\Gamma)}_{\mbox{\scriptsize log.}}=128\pi^2\exp(-4),$$ 
are used to improve the logarithmic accuracy. These coefficients arise from 
asymptotic expansions of elliptic integrals expressing the kinetic energy of 
a flow with a line source and with a vortex string on the ring center-line.

\begin{figure}
\begin{center}
 \epsfig{file=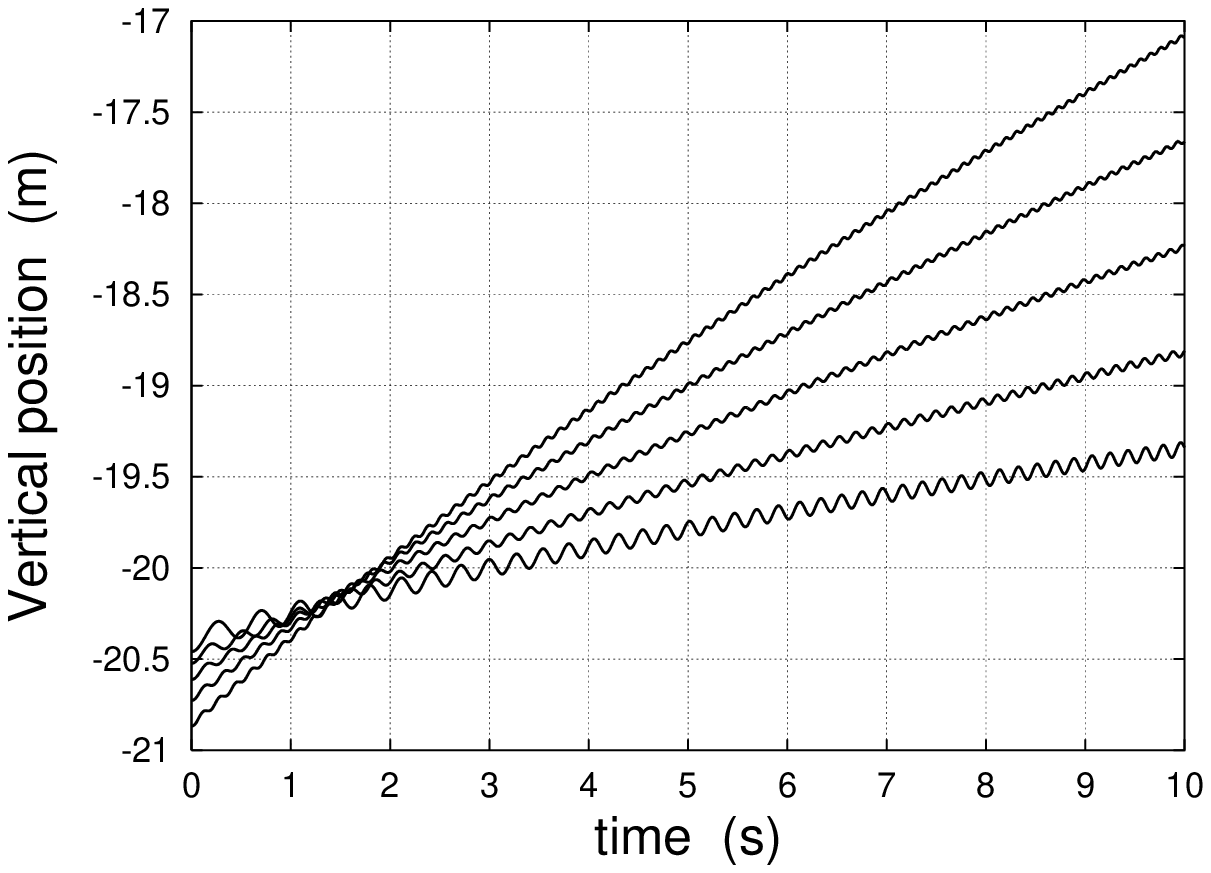,width=80mm}
 \epsfig{file=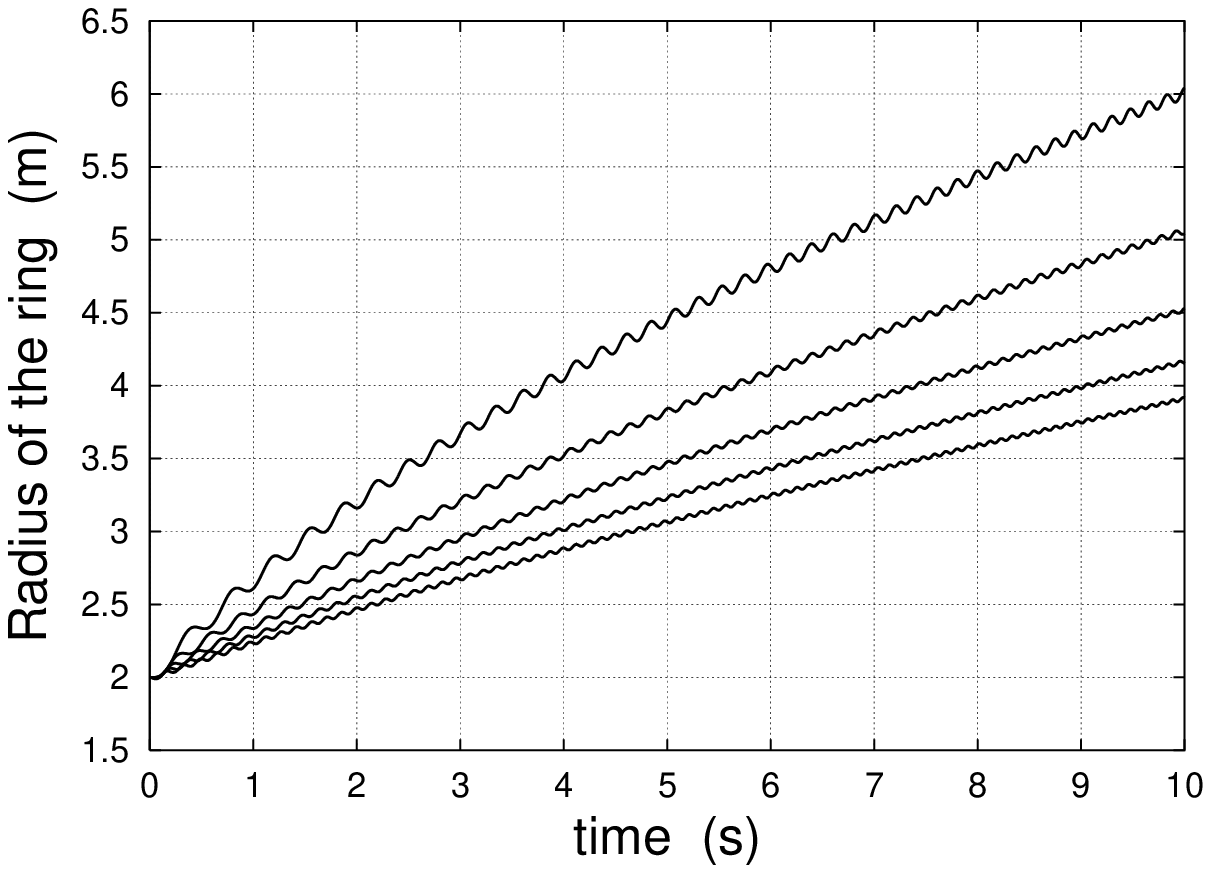,width=80mm} 
 \epsfig{file=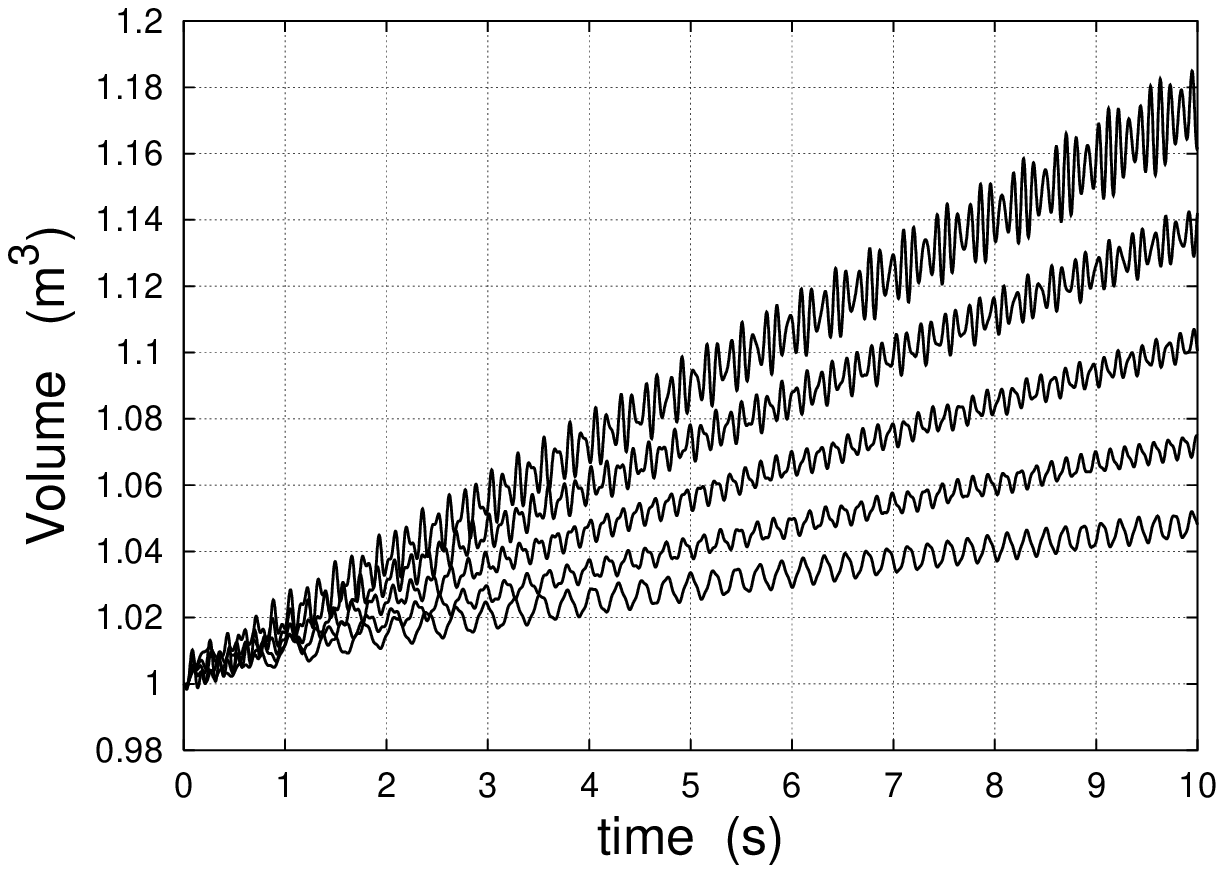,width=80mm}
\end{center}
\caption{\small The evolution of the vertical position $Z(t)$, 
the radius $R(t)$, and the volume ${\cal V}(t)$ of a vortex ring bubble, 
for different values of the circulation, 
$\Gamma=1.0$, $1.5$, $2.0$, $2.5$, $3.0$ $\mbox{m}^2/\mbox{s}$.   
In these simulations the initial data were 
$\dot{\cal V}_0=0$, $\dot R_0=0$, $\dot Z_0=0$,
${\cal V}_0=1.0\,\mbox{m}^3$, $R_0=2.0\,\mbox{m}$, 
and $Z_0=-({\Gamma^2 R_0}/({4{\cal V}_0})+P_0
-2^{1/2}\pi\sigma R_0^{1/2}{\cal V}_0^{-1/2})/g$, 
with $g=9.8$ $\mbox{m}/\mbox{s}^2$, 
$\sigma=7.5\cdot10^{-5}$ $\mbox{m}^3/\mbox{s}^2$.
The parameter $P_0=200$ $\mbox{m}^2/\mbox{s}^2$ approximately corresponds 
to the initial pressure 2 atm.
The curve $Z(t)$ with the largest displacement, the curve $R(t)$ with
the weakest expansion, and the curve ${\cal V}(t)$ that reaches 
the largest values for ${\cal V}$, correspond to the highest value 
of the circulation.} 
\label{ZRV}
\end{figure}

Once the dependence $P({\cal V})$ is given explicitly, 
it is easy to write down and solve numerically the equations of motion 
determined by the Lagrangian (\ref{LagrangianZRV}). 
We used an approximate equation of state for the gas:
$$P({\cal V})=P_0({\cal V}_0/{\cal V})^{1.4}.$$  
Examples of the solutions $Z(t)$, $R(t)$, and ${\cal V}(t)$, 
for several values of the circulation, are presented in Fig.\ref{ZRV}.
The solutions have an oscillating behavior with a drift.
With such initial data, the amplitudes of oscillations for $Z(t)$ and $R(t)$ 
are smaller at larger $\Gamma$. However, oscillations for ${\cal V}(t)$ 
becomes stronger at highest $\Gamma$, that in practice should result in
intensive sound irradiation (this naturally cannot be described
by the present theory of incompressible flows, but needs compressibility).

The system (\ref{LagrangianZRV}) has 6D phase space, but its weakly 
oscillating solutions approximately correspond to the 2D $(R,Z)$-system 
that is obtained by neglecting the inertial terms (quadratic on the time 
derivatives) in Eq.(\ref{LagrangianZRV}) and minimizing the expression 
$\tilde\Pi(R,Z,{\cal V})$ over ${\cal V}$,
$$
\tilde\Pi= \frac{\Gamma^2 R}{4}
\ln\Big(C^{(\Gamma)}_{\mbox{\scriptsize log.}}\frac{R^3}{\cal V}\Big)
+{\cal E}({\cal V})- g Z {\cal V}+2^{3/2}\pi\sigma R^{1/2}{\cal V}^{1/2}.
$$
Thus, we have to solve the equation
$\partial\tilde\Pi(R,Z,{\cal V})/\partial{\cal V}=0$,
\begin{equation}\label{equilibrium}
-\frac{\Gamma^2 R}{4{\cal V}}-P({\cal V})-gZ+
2^{1/2}\pi\sigma R^{1/2}{\cal V}^{-1/2}=0,
\end{equation}
and find from here an equilibrium value ${\cal V}_*(R,Z)$. 
The slow dynamics is approximately described by the Lagrangian
\begin{equation}\label{L_RZ}
\tilde L_{RZ}=-2\pi\Gamma Z R\dot R-\tilde\Pi(R,Z,{\cal V}_*(R,Z)).
\end{equation}
However, since Eq.(\ref{equilibrium}) is in fact already solved for
$Z(R,{\cal V})$, it will be convenient to rewrite the Lagrangian (\ref{L_RZ})
in terms of $R$ and ${\cal V}$:
\begin{eqnarray}
&&\tilde L_{R{\cal V}}=\frac{2\pi\Gamma}{g}
\Big(\frac{\Gamma^2 R}{4{\cal V}}+P({\cal V})-
2^{1/2}\pi\sigma R^{1/2}{\cal V}^{-1/2}\Big)R\dot R\nonumber\\
&&\qquad\quad-\frac{\Gamma^2 R}{4}
\ln\Big(C^{(\Gamma)}_{\mbox{\scriptsize log.}}\frac{R^3}{\cal V}\Big)
-{\cal E}({\cal V})-
2^{3/2}\pi\sigma R^{1/2}{\cal V}^{1/2}\nonumber\\
&&\qquad\quad-{\cal V}\Big(\frac{\Gamma^2 R}{4{\cal V}}+P({\cal V})-
2^{1/2}\pi\sigma R^{1/2}{\cal V}^{-1/2}\Big).
\label{L_RV}
\end{eqnarray}
This dynamical problem is, of course, completely integrable. 
Phase trajectories in the $(R,{\cal V})$-plane are the level contours 
of the effective Hamiltonian
\begin{eqnarray}
&&\tilde H_{R{\cal V}}=
\frac{\Gamma^2 R}{4}\Big(\ln
\Big(C^{(\Gamma)}_{\mbox{\scriptsize log.}}\frac{R^3}{\cal V}\Big)+1\Big)
+{\cal E}({\cal V}) -{\cal V} {\cal E}'({\cal V})\nonumber\\
&&\qquad\qquad+2^{1/2}\pi\sigma R^{1/2}{\cal V}^{1/2}=\tilde E_0=\mbox{const}.
\label{H_RV}
\end{eqnarray}

\begin{figure}
\begin{center}
  \epsfig{file=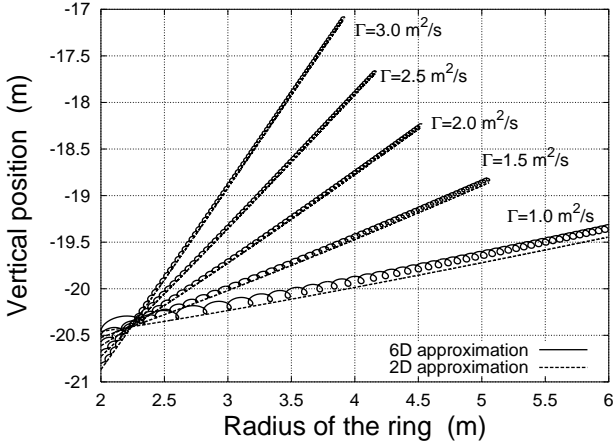,width=85mm}
\end{center}
\caption{\small Comparisons of the bubble ring trajectories in $ZR$-plane,
corresponding to the approximations (\ref{LagrangianZRV}) and (\ref{L_RZ}), 
for different values of the circulation. Parameters as in Fig.\ref{ZRV}.} 
\label{ZR}
\end{figure}

\begin{figure}
\begin{center}
  \epsfig{file=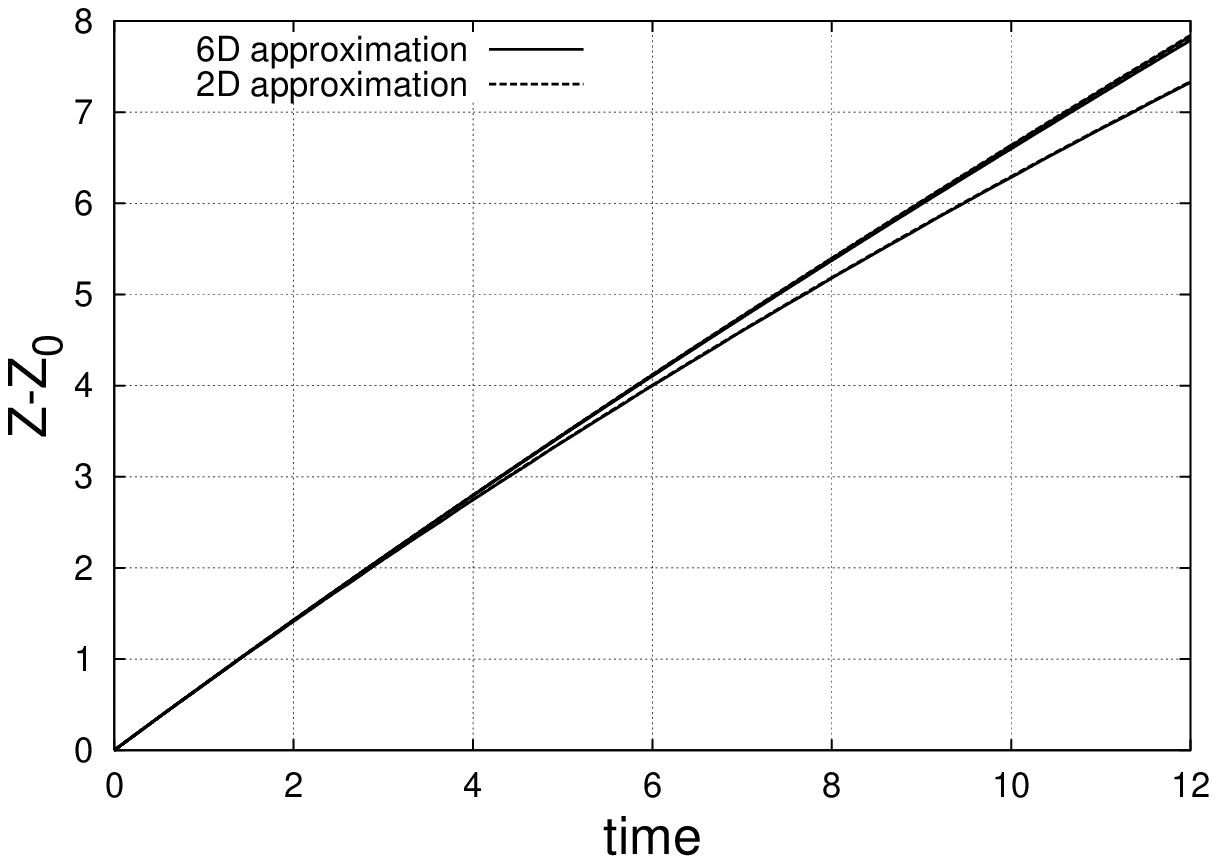,width=72mm}
  \epsfig{file=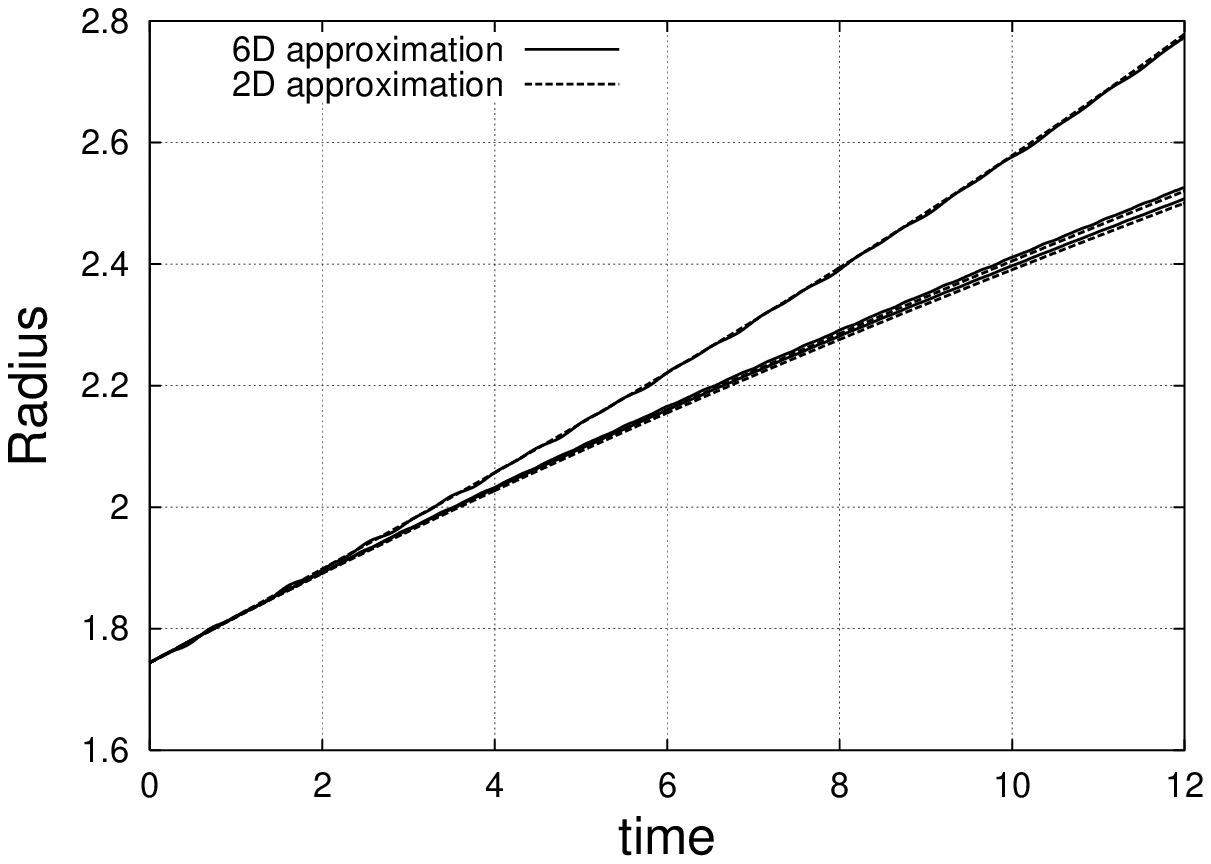,width=72mm}
  \epsfig{file=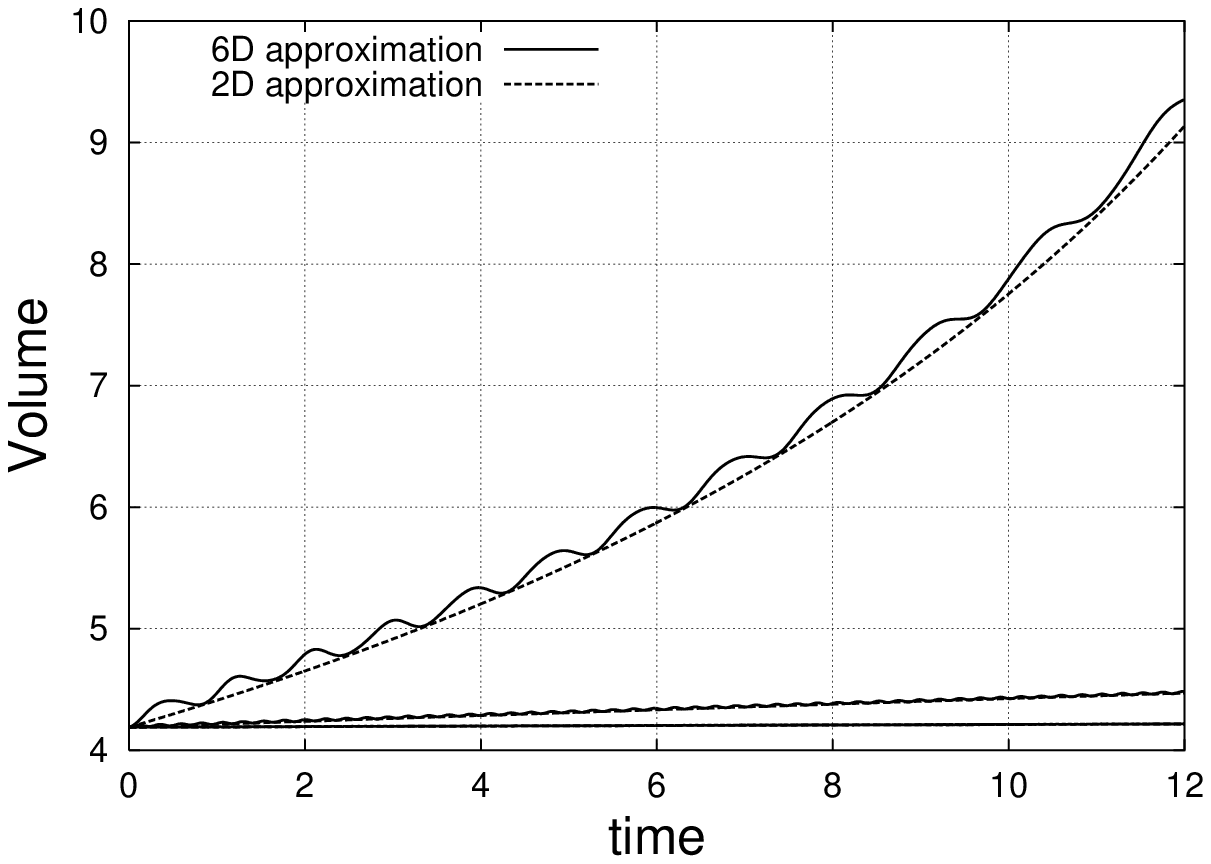,width=72mm}
  \epsfig{file=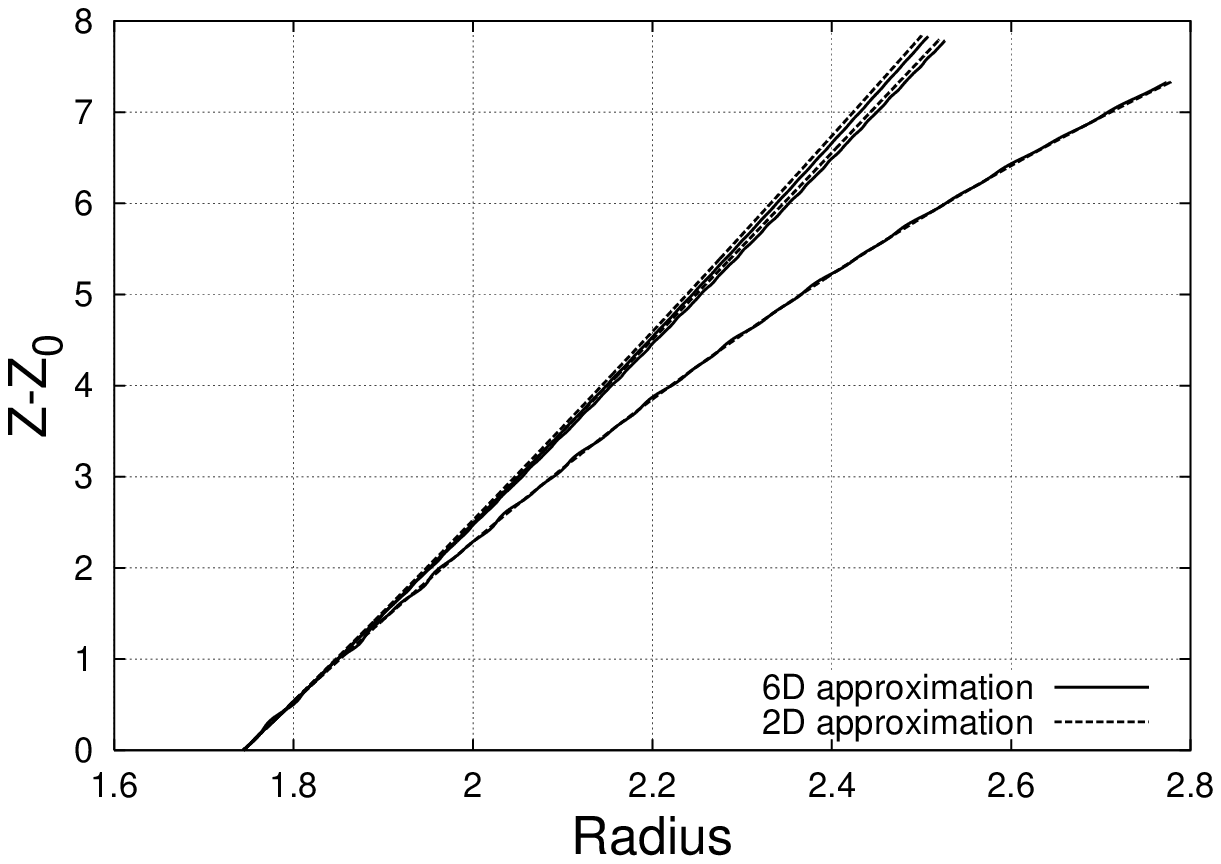,width=72mm}
\end{center}
\caption{\small Simulations for dimensionless quantities: 
$g=1$, $\sigma=0.025$, $\Gamma=5$, ${\cal V}_0=4\pi/3$, 
$R_0/a_0=5$, and $P_0=10,100,1000$. The parameters 
$\dot{\cal V}_0$, $\dot R_0$, $\dot Z_0$ for the 6D approximation
were calculated from the 2D approximation in order to minimize the 
oscillations. The difference between 2D and 6D approximations is almost 
invisible for $Z(t)$ and $R(t)$. The trajectories with large $P_0$ 
are in good agreement with Fig.3 in Ref.\cite{LM1991}.} 
\label{comparison_with_L&M}
\end{figure}

The comparison between the drifting solutions of the 6D system 
(\ref{LagrangianZRV}) and the corresponding solutions of the 2D system 
(\ref{L_RZ}) is given in Fig. \ref{ZR}.
We observe that the agreement is very good in particular for the high 
values of the circulation.
Furthermore, the overall evolution is qualitatively in agreement with 
the numerical results of Lundgren and Mansour \cite{LM1991}, 
especially for large values of the parameter $P_0$, 
when the relative variation of the volume is small 
(see Fig.\ref{comparison_with_L&M}, where all the plotted quantities 
have been made dimensionless as in Ref.\cite{LM1991}).

It should be kept in mind that applicability of the axisymmetric 
(both 6D and 2D) models requires, besides $2\pi^2R^3\gg{\cal V}$ and 
$\Gamma^2\gg g({\cal V}/R)^{3/2}$, at least one more condition:
\begin{equation}\label{sausage_stability}
R^{-1/2}{\cal V}^{1/2}\le\frac{\Gamma^2}{2^{3/2}\pi\sigma}.
\end{equation}
This inequality makes constant-cross-area configurations be
stable to longitudinal sausage-like perturbations \cite{Ponstein1959}.
If a vortex ring bubble violates this criterion in the course of motion, 
then it will be destabilized by surface tension and the development 
of the instability, together with viscous effects, will result in 
transformation of the ring into a closed chain of smaller bubbles
connected by thin vortex filaments, as it is observed experimentally
(concerning this phenomenon, Lundgren and Mansour \cite{LM1991}
refer to a private communication by D. McSweeny (1990)).
However, we found with realistic dependences ${\cal E}({\cal V})$
that the Hamiltonian evolution of the spreading ring, described above, 
typically moves away from the instability criterion. 
Thus, to destroy the ring, viscous diffusion of the vorticity seems 
to be necessary \cite{Pedley,LM1991}. This will result 
in a slow decrease of the circulation along a contour
just above the bubble surface. Ultimately, this circulation becomes 
too small to satisfy the stability condition (\ref{sausage_stability}),
the instability will then evolve and the ring will tend to break up.


\section{Summary}

In this work we have developed a variational approach for the theoretical
study of the ideal incompressible irrotational flows with a distorted
toroidal bubble, in the case when the velocity potential is a multi-valued
function.
Using this method, we have derived exact pseudo-differential equations of 
motion for purely two-dimensional flows with circulation around a single
cavity. Also we have suggested a simplified Lagrangian for a three-dimensional
thin hollow vortex tube. As a simple particular case, the axisymmetric
vertical motion of a spreading vortex ring bubble with a compressed gas 
inside has been considered. Approximate solutions of a corresponding
finite-dimensional dynamical system have been obtained.

\subsection*{Acknowledgments}
These investigations were supported by INTAS (grant No. 00-00292).
The work of V. R. also was supported by RFBR and 
by the Russian State Program of Support of the Leading Scientific Schools.

\appendix

\section{Variation of $\int{\cal L}_{\mbox {\scriptsize conf.}} dt$}

Here we extract the equations of motion (\ref{dot_z}) and (\ref{dot_psi})
from the expression (\ref{L_z_psi}).
First of all, variation of the action $\int{\cal L}_{\mbox {\scriptsize conf.}} dt$
by $\delta\psi$ gives the kinematic condition in terms of the conformal 
variables
$$
\frac{(\dot \zeta\bar \zeta'-\dot{\bar \zeta}\zeta')}{2i}=\hat M \psi.
$$
Now a standard procedure is to divide this equation by $|\zeta'|^2$,
$$
\frac{\dot \zeta}{\zeta'}-\frac{\dot {\bar \zeta}}{\bar \zeta'}
=\frac{2i\hat M \psi}{|\zeta'|^2},
$$
and apply the operator $\hat P^{(-)}$ which excludes Fourier-harmonics with
positive $m$. As the result, we get Eq.(\ref{dot_z})
$$
\dot \zeta=\zeta'\hat P^{(-)}\left(\frac{2i\hat M \psi}{|\zeta'|^2}\right).
$$

The variation of $\int{\cal L}_{\mbox {\scriptsize conf.}} dt$ 
by $\delta\bar \zeta$ and subsequent
exlusion of the Lagrangian multiplier $\lambda$ gives the equation
\begin{eqnarray}
\hat P^{(-)}\Big(\Big(-\dot \zeta(\gamma+\psi')+
\Big(g\frac{(\zeta-\bar \zeta)}{2i}+\dot\psi-\frac{\gamma^2}{2r^2}\Big)\zeta'
&&\nonumber\\
+\frac{\gamma^2}{2}\cdot\frac{ie^{i\vartheta}}
{\int \bar \zeta(\vartheta)e^{i\vartheta}\frac{d\vartheta}{2\pi}}\Big)
e^{-i\vartheta}\Big)=0.&&\nonumber
\end{eqnarray}

The statement $\hat P^{(-)}f=0$ means that $f$ contains only
harmonics with positive $m$. The function $\bar \zeta'e^{i\vartheta}$ 
does not contain harmonics with negative $m$. Therefore the equality
$\hat P^{(-)}(f\bar \zeta'e^{i\vartheta})=0$ is true. 
In our case this results in
\begin{eqnarray}
\hat P^{(-)}\Big(-\dot \zeta \bar \zeta'(\gamma+\psi')+
\Big(g\frac{(\zeta-\bar \zeta)}{2i}+\dot\psi-\frac{\gamma^2}{2r^2}\Big) 
|\zeta'|^2
&&\nonumber\\
\label{minus}
+\frac{\gamma^2}{2}\cdot
\frac{i\bar \zeta'e^{i\vartheta}}{\bar \zeta_1}\Big)=0.
\end{eqnarray}
It is easy to check that
$$
\hat P^{(-)}\left(\frac{\gamma^2}{2}\cdot\frac{i\bar \zeta'e^{i\vartheta}}
{\bar \zeta_1}\right)=\frac{\gamma^2}{4}.
$$
Now we have to take the real part of Eq.(\ref{minus}). 
Using the property $\hat P^{(-)}+\hat P^{(+)}=1$,
we get the equation which in fact is solved for $\dot\psi$,
\begin{eqnarray}
&&\Big(g\frac{(\zeta-\bar \zeta)}{2i}+\dot\psi-\frac{\gamma^2}{2r^2}\Big)
|\zeta'|^2=
\nonumber\\
&&\left(\hat P^{(-)}\left\{
(\gamma+\psi')|\zeta'|^2\hat P^{(-)}\left(\frac{2i\hat M \psi}
{|\zeta'|^2}\right)
\right\}
+c.c.\right) -\frac{\gamma^2}{2}.\nonumber
\end{eqnarray}
After simplification that uses the equality
$$
\hat H(2\psi'\hat M \psi)=(\psi')^2-(\hat M\psi)^2,
$$
we  finally obtain Eq.(\ref{dot_psi}).

\section{Expression (\ref{psi_pq})}

To solve for $\psi$ the equation
$$
p=\gamma\hat H q-\hat P^{(-)}[\psi(\beta-i\beta')]
-\hat P^{(+)}[\psi(\bar \beta+i\bar \beta')],
$$
we perform the following steps. First, let us separate
the harmonics with $m\le 0$:
\[
\hat P^{(-)}[\psi(\beta-i\beta')+p-\gamma\hat H q]
=i\alpha=2i\hat P^{(-)}\alpha,
\]
where $\alpha$ is an unknown real quantity constant on $\vartheta$. 
Now we use the same trick as in the Appendix A, 
i.e., we multiply the above equation by the function $\bar \beta+i\bar \beta'$,
\[
\hat P^{(-)}[\psi|\beta-i\beta'|^2+
(p-\gamma\hat H q)(\bar \beta+i\bar \beta')]
\]
\[
=2i\alpha\hat P^{(-)}(\bar \beta+i\bar \beta')=i\alpha q_0,
\]
and take the real part,
\[
-\psi|\beta-i\beta'|^2
\]
\[
=\hat P^{(-)}[(p-\gamma\hat H q)(\bar \beta+i\bar \beta')]
+\hat P^{(+)}[(p-\gamma\hat H q)(     \beta-i     \beta')].
\]
Then we simplify it and use the explicit formula 
$$
\beta-i \beta'=(q-\hat M q)+i\hat H(q-\hat M q),
$$
that results in the expression (\ref{psi_pq}).


\begin{thebibliography}{99}

\bibitem{WD63} J.K. Walters and J.F. Davidson,
J. Fluid Mech. {\bf 17}, 321 (1963).


\bibitem{LM1991} T. S. Lundgren and N. N. Mansour,
J. Fluid Mech. {\bf 224}, 177 (1991).

\bibitem{SussmanSmereka} M. Sussman and P. Smereka,
J. Fluid Mech. {\bf 341}, 269 (1997).

\bibitem{bubblerings}  http://www.bubblerings.com/bubblerings/

\bibitem{Ponstein1959} J. Ponstein, Appl. Sci. Res. A {\bf 8}, 425 (1959).

\bibitem{Pedley} T. J. Pedley, J. Fluid Mech. {\bf 30}, 127 (1967);
T. J. Pedley, J. Fluid Mech. {\bf 32}, 97 (1968).

\bibitem{Arnold}  V. I. Arnol'd, {\it Mathematical Methods of Classical
Mechanics}, 2nd edition (Springer-Verlag, New York, 1989).

\bibitem{Salmon} R. Salmon, Ann. Rev. Fluid Mech. {\bf 20}, 225 (1988).

\bibitem{ZK97} V. E. Zakharov and E. A. Kuznetsov,
Usp. Fiz. Nauk {\bf 167}, 1037 (1997) [Phys. Usp. {\bf 40}, 1087 (1997)].

\bibitem{Morrison98} P. J. Morrison, Rev. Mod. Phys. {\bf 70}, 467 (1998).

\bibitem{IL} V.I. Il'gisonis and V.P. Lakhin,
Plasma Phys. Rep. {\bf 25}, 58 (1999).

\bibitem{HMR} D.D. Holm, J.E. Marsden, and T.S. Ratiu,
arXiv:chao-dyn/9801015; arXiv:chao-dyn/9903035.

\bibitem{Z68} V. E. Zakharov, Prikl. Mekh. Tekh. Fiz. {\bf 2}, 86 (1968).

\bibitem{LMMR86} D. Lewis, J. Marsden, R. Montgomery and T. Ratiu,
Physica D {\bf 18}, 391 (1986).

\bibitem{DKSZ96} A. I. Dyachenko, E. A. Kuznetsov, M. D. Spector,
and V. E. Zakharov, Phys. Lett. A {\bf 221}, 73 (1996).

\bibitem{DLZ95} A. I. Dyachenko, Y. V. L'vov, and V. E. Zakharov,
Physica D {\bf 87}, 233 (1995).

\bibitem{ZD96} V. E. Zakharov and A. I. Dyachenko, 
Physica D {\bf98}, 652 (1996).

\bibitem{DZK96} A. I. Dyachenko, V. E. Zakharov, and E. A. Kuznetsov,
Fiz. Plazmy {\bf 22}, 916 (1996) [Plasma Phys. Rep. {\bf 22}, 829 (1996)].
    
\bibitem{D2001} A. I. Dyachenko, Doklady Akademii Nauk {\bf 376}, 27 (2001) 
[Doklady Mathematics {\bf 63}, 115 (2001)].

\bibitem{ZDV2002} V. E. Zakharov, A. I. Dyachenko, and O. A. Vasilyev,
European Journal of Mechanics B/Fluids {\bf 21}, 283 (2002).


\bibitem{LL6} L. D. Landau and E. M. Lifshitz, {\it Fluid Mechanics}
(Pergamon Press, New York, 1987).

\bibitem{HBGL1998} S. Hilgenfeldt, M. P. Brenner, S. Grossmann, and D. Lohse,
J. Fluid Mech. {\bf 365}, 171 (1998).

\bibitem{R2001PRE} V. P. Ruban, Phys. Rev. E {\bf 64}, 036305 (2001).

\bibitem{R2000PRD} V. P. Ruban, Phys. Rev. D {\bf 62}, 127504 (2000).

\bibitem{Hasimoto} H. Hasimoto, J. Fluid Mech. {\bf 51}, 477 (1972).



\end{thebibliography}
\end{document}